\documentclass[sigconf]{acmart}

\usepackage{multirow}
\usepackage{todonotes}
\usepackage{todonotes}
\usepackage{enumitem}
\usepackage{hyperref}
\usepackage{graphicx}
\usepackage{subcaption}
\usepackage{amsmath}

\AtBeginDocument{%
  \providecommand\BibTeX{{%
    \normalfont B\kern-0.5em{\scshape i\kern-0.25em b}\kern-0.8em\TeX}}}

\copyrightyear{2024}
\acmYear{2024}
\setcopyright{rightsretained}
\acmConference[ICAIF '24]{5th ACM International Conference on AI in Finance}{November 14--17, 2024}{Brooklyn, NY, USA}
\acmBooktitle{5th ACM International Conference on AI in Finance (ICAIF '24), November 14--17, 2024, Brooklyn, NY, USA}
\acmDOI{10.1145/3677052.3698695}
\acmISBN{979-8-4007-1081-0/24/11}

\begin{document}

\title{ARL-Based Multi-Action Market Making with Hawkes Processes and Variable Volatility}

\author{Ziyi Wang}
\email{ziyi.1.wang@kcl.ac.uk}
\affiliation{%
  \institution{King's College London}
  \city{London}
  \country{United Kingdom}
}

\author{Carmine Ventre}
\email{carmine.ventre@kcl.ac.uk}
\affiliation{%
  \institution{King's College London}
  \city{London}
  \country{United Kingdom}
\email{carmine.ventre@kcl.ac.uk}
}
\author{Maria Polukarov}
\email{maria.polukarov@kcl.ac.uk}
\affiliation{%
  \institution{King's College London}
  \city{London}
  \country{United Kingdom}
\email{maria.polukarov@kcl.ac.uk}
}

\renewcommand{\shortauthors}{Wang et al.}

\begin{abstract}
We advance market-making strategies by integrating Adversarial Reinforcement Learning (ARL), Hawkes Processes, and variable volatility levels while also expanding the action space available to market makers (MMs). To enhance the adaptability and robustness of these strategies -- which can quote always, quote only on one side of the market or not quote at all -- we shift from the commonly used Poisson process to the Hawkes process, which better captures real market dynamics and self-exciting behaviors. We then train and evaluate strategies under volatility levels of 2 and 200. Our findings show that the 4-action MM trained in a low-volatility environment 
effectively adapts to high-volatility conditions, maintaining stable performance and providing two-sided quotes at least 92\% of the time. This indicates that incorporating flexible quoting mechanisms and realistic market simulations significantly enhances the effectiveness of market-making strategies.
\end{abstract}



\maketitle

\section{Introduction}
Market makers (MMs) are integral to financial markets, enhancing liquidity and stability by consistently providing bid and ask quotes, which promotes efficient and orderly trading. While they derive profits from the bid-ask spread, they face significant challenges due to the complexity of market environments and price volatility. Recent research has focused extensively on risk management and strategy optimization for market makers. For instance, \cite{spooner2020robust} demonstrates that adversarial reinforcement learning (ARL) can produce market-making agents that exhibit strong robustness under epistemic uncertainty. By dynamically adjusting parameters in market dynamics to create various adversarial environments, they provide valuable insights into developing market-making strategies that are adaptable to complex and fluctuating trading conditions. Building on this, \cite{wang2023robust} extends the action space of market makers by relaxing the assumption that they must always quote at each time step. This flexibility allows market makers to dynamically choose among providing two-sided quotes, offering only a bid price, providing only an ask price, or declining to quote. Their exploration of flexible quoting mechanisms reveals how they can enhance market-maker profitability and risk management by leveraging market rules more effectively.

In this study, to enhance both the realism of market dynamics simulation and the adaptability of market making strategies, we build upon the previous approach of training multi-action space market-making agents in adversarial environments. We introduce two significant contributions. Firstly, we transition from using a Poisson process to a Hawkes process for modeling execution dynamics. The Hawkes process, with its self-exciting feature, has the potential to capture price jumps and self-reinforcing trading behaviors observed in real markets, which may enhance the alignment of our market model with actual market conditions. Secondly, we investigate the effects of volatility on market maker strategies by employing two distinct volatility levels: low and high. This range allows us to explore how variations in price dynamics impact strategy performance. Although increasing volatility introduces greater market complexity and risk, it enables a more accurate evaluation and optimization of the market maker strategies under real-world conditions. These enhancements aim to more effectively simulate market instability and provoke extreme scenarios, thus improving the robustness of the market-making strategies.

\begin{figure*}
    \centering
    \includegraphics[width=\textwidth]{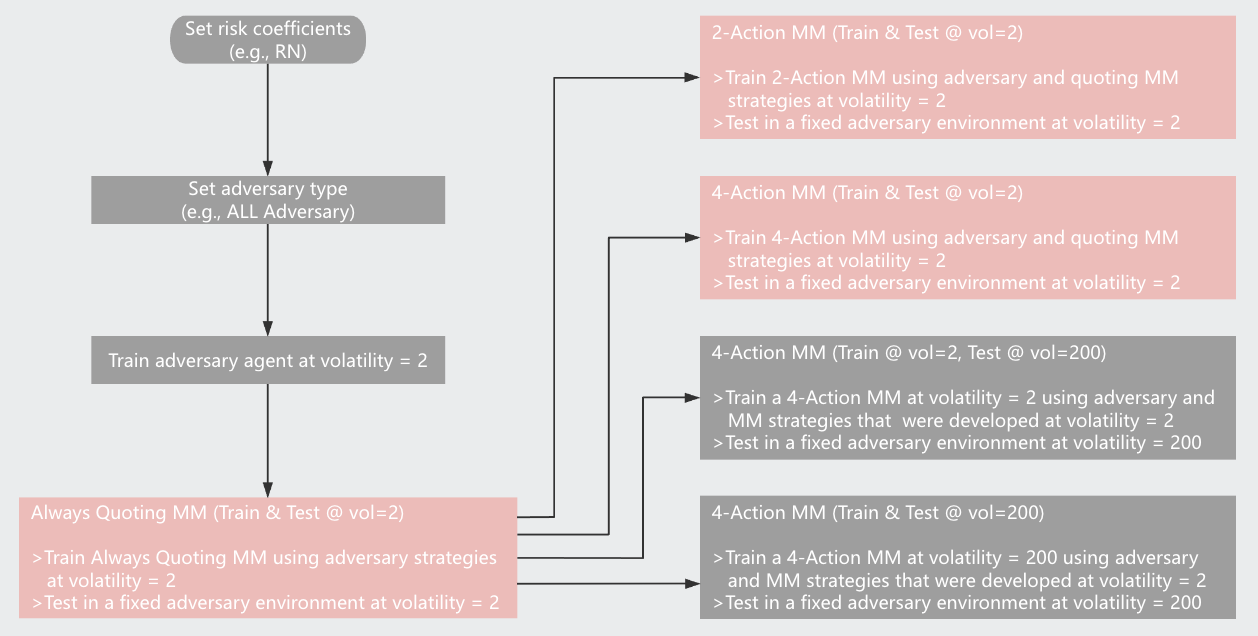}
  \caption{Training and Testing Multi-Action Market Makers under Different Volatility Levels}
  \label{process}
\end{figure*}

As illustrated in Figure \ref{process}, we investigate the impact of volatility on market maker performance using the 4-Action MM strategies trained and tested under varying volatility levels. The results show that the 4-Action MM strategy trained in a low-volatility environment but tested in a high-volatility environment still demonstrates relatively stable performance, indicating effective adaptation to the new conditions. This shows that, within our experimental setup, multi-action space market-making agents trained in adversarial environments with the Hawkes process are effectively capable of adapting to significant changes in market volatility. Moreover, the 4-Action MM strategy trained and tested in a high-volatility environment -- but utilizing adversarial agents and quoting strategies developed in a low-volatility setting -- tends to exhibit more conservative quoting behavior, such as opting not to quote or making unilateral quotes to mitigate the risks associated with high volatility. This behavior highlights the potential motivation for market makers to use flexible quoting mechanisms and avoid risks when facing unfavorable conditions.

\section{Related Work}
\vspace{1ex} \noindent {\bf Stochastic Control Approach.}
Market-making strategies have significantly evolved over time, driven by both theoretical advancements and practical applications. Initial research focused on optimizing bid and ask prices using stochastic dynamic programming. \citet{ho1981optimal} laid the groundwork in this area, developing dynamic programming methods to determine optimal pricing strategies that maximize the expected utility of a market maker's wealth. Building on this, \citet{avellaneda2008high} introduced a two-step method that incorporated limit order book dynamics and inventory risk management, offering a more nuanced understanding of market-making under complex conditions. Subsequent work by \citet{gueant2013dealing} addressed inventory constraints in market-making models, and \citet{fodra2012high} refined these models by integrating risk-aversion parameters and extending their application to various mid-price processes.

Despite these advancements, many theoretical models rely on the assumption of perfect knowledge of market dynamics, which is rarely applicable in real-world scenarios. \citet{cartea2015algorithmic} pointed out that inaccuracies in model specification can undermine optimal strategy selection. To address these challenges, \citet{gueant2017optimal} investigated strategies under model uncertainty, focusing on how market makers can adapt to imperfect information by adjusting quotes to manage inventory risk and adverse selection.

\vspace{1ex} \noindent {\bf Reinforcement Learning Approach.} 
The application of reinforcement learning (RL) has introduced new approaches to market-making.  \citet{chan2001electronic} were early adopters of RL in this field, evaluating various RL algorithms and finding that actor-critic methods were particularly effective in complex environments. \citet{abernethy2013adaptive} developed strategies that minimize quote spreads using online learning algorithms. \citet{spooner2018market} advanced these methods by creating a high-fidelity limit order book market simulation and using high-frequency data to train inventory-sensitive market makers. Building on these contributions, \citet{spooner2020robust} employed adversarial reinforcement learning (ARL) to develop robust trading strategies capable of handling model uncertainty. Their approach, which adapts single-agent models into zero-sum games with adversarial training.

\citet{wang2023robust} further extend these advancements by exploring flexible quoting mechanisms. By relaxing the requirement for market makers to quote at every time step, Wang introduces a model where market makers can choose among providing two-sided quotes, offering only a bid price, providing only an ask price, or opting not to quote. This extension allows for greater flexibility in market-making strategies and has the potential to enhance risk management and profitability by adapting more effectively to market conditions. Additionally, \citet{jerome2023mbt} have introduced the \texttt{mbt\_gym} Python module, which provides a suite of gym environments for training RL agents in model-based trading problems within limit order books. This module supports efficient, vectorized implementations that accelerate RL training and offers a flexible framework for combining various model aspects. This work motivates the use of advanced stochastic processes, such as Hawkes Processes, for capturing complex market dynamics and self-exciting behaviors in trading models.

\vspace{1ex} \noindent {\bf Hawkes Processes in Modeling Market Dynamics.} 
Hawkes processes, introduced by \citet{hawkes1971spectra}, have become essential in modeling complex financial dynamics. Early applications, such as those by \citet{hewlett2006clustering}, utilized bivariate Hawkes processes to predict trade price impacts in foreign exchange markets, aiding in optimal liquidation strategies. \citet{large2007measuring} conducted an early study applying Hawkes processes to order book modeling. Additionally, \citet{toke2012modelling} demonstrated that a simple bivariate Hawkes process effectively models trades-through in limit order books.

The relevance of Hawkes processes in understanding trading behaviors in financial markets is further supported by the work of \cite{bacry2013modelling}, \cite{bacry2014hawkes}, and \cite{bacry2016estimation}. In particular, \citet{bacry2015hawkes} provided a comprehensive overview of Hawkes processes, emphasizing their flexibility in addressing various high-frequency finance issues, such as estimating volatility, market stability, and optimal execution strategies. 

Building on previous empirical and numerical studies, \citet{abergel2015long} introduced a mathematical model of limit order books based on Hawkes processes with exponential kernels. Furthermore, the authors of \cite{lu2017limit,lu2018high} effectively modeled limit order book (LOB) dynamics using high-dimensional Hawkes processes, enhancing the understanding of order flow within the LOB.

Recent research \cite{swishchuk2020general,guo2020multivariate1,guo2020multivariate2} focused on general compound Hawkes processes. Building on this, \citet{roldan2022optimal} addressed optimal control problems in high-frequency trading, including optimal acquisition, liquidation, and market making.

Overall, these research findings not only enrich the theoretical framework of Hawkes processes but also illustrate their effectiveness in modeling complex interactions within limit order books, contributing to a deeper understanding of trading and market-making strategies.

Our research builds on these prior contributions by integrating Adversarial Reinforcement Learning with Hawkes Processes and varying volatility levels. By moving beyond the commonly used Poisson process, we aim to better capture realistic market dynamics and self-exciting behaviors. Our results demonstrate that a four-action market maker, trained in a low-volatility environment, can adapt effectively to high-volatility conditions while maintaining consistent performance. This underscores the importance of flexible quoting mechanisms and realistic market simulations in refining market-making strategies.

\section{Problem Description}
\subsection{Market Dynamics}
\subsubsection{Price Dynamics} 
 Our model draws inspiration from the market-making framework proposed by \cite{avellaneda2008high}. The price dynamics of the asset follow a stochastic process modeled as a Brownian motion with drift. Specifically,

\begin{equation*}
Z_{n+1}=Z_n+b_n\Delta t+\sigma_n W_n,
\end{equation*}

where $Z_n$ is the asset price at time $t=n$, $b_n$ is the drift coefficient, $\sigma_n$ is the volatility coefficient, and $W_n$ is an independent, normally-distributed random variable from a normal distribution $N(0,\Delta t)$ with mean $0$ and variance $\Delta t$ in any finite time interval $\Delta t$ (i.e., the variance increases linearly with the length of the time interval). For our experiments, we set the initial price $Z_0=100$ and fixed the volatility $\sigma_n=2$ for all $n$.

For an asset, the bid offset $\delta_n^+$ represents the difference between the market maker's expected buy (bid) price, $P_n^+$, and the current market price $Z_n$ of the asset. Similarly, the ask offset $\delta_n^-$ is the difference between the market maker's expected sell (ask) price, $P_n^-$, and the current market price. These offsets can be mathematically expressed as:

\begin{equation*}
\delta_n^\pm = \pm (Z_n - P_n^\pm) \ge 0.
\end{equation*}

\subsubsection{Execution Dynamics}

At each time step, the probability of executing a trade between the market maker and the environment is determined using a Hawkes process. This approach accounts for both the current market conditions and the market maker's inventory. The trade arrival intensity is given by:

\begin{equation*}
    \lambda_n^\pm = \text{Hawkes\_intensity} \cdot e^{-k_n^\pm \delta_n^\pm}.
\end{equation*}

Here, \texttt{Hawkes\_intensity} is updated dynamically based on past trading activity and mean-reversion factors. The parameter \(k_n^\pm\), where \(k_n^\pm > 0\), describes the distribution of volume in the order book.

This approach is inspired by the \texttt{HawkesArrivalModel} class in \cite{jerome2023mbt}, which utilizes the Hawkes process to model dynamic environments for market-making problems, and it also incorporates the parameters of the Hawkes process defined therein.

In the \texttt{HawkesArrivalModel} class, the \texttt{Hawkes\_intensity} is computed as follows:

\begin{align}
\text{Hawkes\_intensity} = & \text{Hawkes\_intensity} + \text{mean\_reversion\_speed} \nonumber \\
& \times (\text{baseline\_arrival\_rate} - \text{Hawkes\_intensity}) \nonumber \\
& \times \text{dt} + \text{jump\_size} \times \text{last\_match\_result}
\end{align}

where:

\begin{itemize}
    \item {mean\_reversion\_speed} (60.0): Controls how quickly the intensity reverts to the baseline.
    \item {baseline\_arrival\_rate} (10.0): The base intensity level towards which the system reverts.
    \item {jump\_size} (40.0): The increase in intensity due to recent trading results.
    \item {dt} (0.005): The time step increment.
    \item {last\_match\_result}: The outcome of the most recent trade, where 1 indicates a successful trade and 0 indicates no trade.
\end{itemize}

\subsubsection{Inventory Dynamics} 
The cumulative return, or wealth, of the market maker is given by the sum of cash and inventory values:

\begin{equation*}
\Pi(X, H, Z) = X + HZ
\end{equation*}

where $X$ denotes cash, calculated as:

\begin{equation*}
\begin{aligned}
X_{n+1} &= X_n + P_n^- \Delta N_n^- - P_n^+ \Delta N_n^+ \\
&= X_n + \delta_n^- \Delta N_n^- + \delta_n^+ \Delta N_n^+ - Z_n \Delta H_n
\end{aligned}
\end{equation*}

Here, $N_n^+$ ($N_n^-$) represents the cumulative number of assets bought (sold) by the market maker at bid (ask) prices up to time $t=n$, and $\Delta N_n^\pm = N_{n+1}^\pm - N_n^\pm$. The cash inflow or outflow from each transaction is determined by the bid/ask price, and the value of inventory changes with the price. The current inventory $H_n$ is given by:

\begin{equation*}
H_n = (N_n^+ - N_n^-) \in [\underline{H}, \overline{H}]
\end{equation*}

where $H_n$ represents the difference between the number of assets bought and sold, constrained within the bounds $[\underline{H}, \overline{H}]$.

\subsection{Market Participants}

In this study, the market maker and adversary engage in a zero-sum game, with the adversary representing other market participants who seek to profit at the expense of the market maker.

\subsubsection{Adversarial Environment}

The adversary influences the market dynamics through three key parameters: $b_n$ (drift coefficient), $A_n^\pm$ (baseline arrival rate in the Hawkes process), and $k_n^\pm$ (distribution of volume). These parameters impact both price movements and trade execution.

\begin{description}[leftmargin=10pt]
\item[Fixed Adversary.] In this scenario, the adversary's parameters are constant throughout all episodes: $b_n=0$, $A_n^\pm=10$, and $k_n^\pm=1.5$. This setup results in a stationary environment, where the market maker's interaction can be modeled as a single-agent problem with fixed dynamics.

\item[Random Adversary.] At the start of each episode, the adversary's parameters are chosen independently and uniformly at random from predefined ranges: $b_n \in [-5, 5]$, $A_n^\pm \in [7.5, 12.5]$, and $k_n^\pm \in [1.125, 1.875]$. Once chosen, these parameters remain fixed for the duration of the episode.

\item[Strategic Adversary.] In this setup, the adversary adjusts the market parameters $b_n$, $A_n^\pm$, and $k_n^\pm$ to maximize its own reward, which is directly opposed to the market maker's reward. This strategic approach entails that the adversary selects parameters to minimize the market maker's performance. Additionally, we explore the impact of individual parameters by setting up strategic adversaries that focus on controlling only one of $b_n$, $A_n^\pm$, or $k_n^\pm$ at a time.
\end{description}

\subsubsection{Types of Market Maker Agents}

The simulated market includes three types of market maker agents, each with distinct action spaces and objectives:

\begin{description}[leftmargin=10pt]

\item[Always Quoting Market Maker Agent.] This agent's goal is to determine the optimal bid and ask offsets based on the prevailing market conditions. The action space for this agent includes two continuous variables: the ask offset and the bid offset, each ranging from 0 to 3.

\item[Market Maker Agent with Two Action Spaces.] This agent decides whether to provide both bid and ask quotes or to refrain from quoting altogether in each time step to manage risk. The action space is discrete, consisting of \(\{0, 1\}\), where \(0\) indicates no quoting and \(1\) signifies quoting both the bid and ask.

\item[Market Maker Agent with Four Action Spaces.] This agent determines whether to provide bid-ask quotes, single-sided quotes, or not quote at all in each time step to mitigate risk. Its action space is \(\{0, 1, 2, 3\}\), where \(0\) means no quoting, \(1\) indicates quoting both the bid and ask, \(2\) represents quoting only the ask, and \(3\) denotes quoting only the bid.
\end{description}

\section{Training Process Overview}
Figure \ref{process} illustrates the training process of market-making agents under adversarial reinforcement learning. In this experiment, we first set the risk coefficients, for example, RN ($\eta = 0.0$ and $\zeta = 0.0$). The reward function for the market maker is defined as:

\begin{equation*}
R_n = \Delta \Pi_n - \zeta H_n^2 - \left\{
\begin{aligned}
&0 &\text{for } t < T, \\ 
&\eta H_n^2 &\text{otherwise.}
\end{aligned}
\right.
\end{equation*}

If $\eta=0$ and $\zeta=0$, then $R_n = \Delta \Pi_n$, which we define as the risk-neutral (RN) scenario. In cases where $\eta$ and $\zeta$ are non-zero, the agent's reward varies with the current inventory $H_n$, representing a risk-averse (RA) scenario. We define the adversary's reward as the opposite of the market maker's, as the adversary and the market maker are modeled in a zero-sum game. Our experiment examines six different sets of RA coefficients.

Next, we set the type of adversary, for instance, ALL Adversary. At this point, we begin the training process for the agents, using reinforcement learning algorithms from the \texttt{tf\_agents} library. Since we define the adversary and the market maker in a zero-sum game, we first train the adversary agent. We use the SAC algorithm, a soft actor-critic method by \cite{haarnoja2018soft}, to train the ALL Adversary agent under volatility=2. The adversary agent is trained for at least 50,000 episodes. The policies update every 1000 time-steps with a learning rate of $3\times{10}^{-4}$ for both the actor and critic policies. The batch size is $64$. The well-trained ALL Adversary agent dynamically adjusts three key market parameters at each time step to maximize its own reward. The action ${a=(b}_n,A_n^\pm,k_n^\pm)$ is given by a vector that contains the drift, Hawkes baseline rate and decay, where $b_n=b\in[-5,5]$, $A_n^\pm=A\in[7.5,12.5]$ and $k_n^\pm=k\in[1.125,1.875]$.

Subsequently, we use the SAC algorithm to train the always-quoting market maker agent under volatility=2, leveraging the trained adversary strategy. The parameter configuration for the training process is the same as above. The agent is then tested in a fixed adversary environment at volatility=2. At this point, we have the trained adversary (Train \& Test @ vol=2) and the Always Quoting MM (Train \& Test @ vol=2). We then proceed to train the multi-action space market-making agents. Using the DQN algorithm, we train the multi-action space market-making agents for at least 50,000 episodes. The policies update every time-step with a learning rate of $1 \times 10^{-4}$. The batch size is $64$.

For example, \textbf{2-Action MM (Train @ vol=2)}: Under volatility=2, using the adversary (Train \& Test @ vol=2) and Always Quoting MM (Train \& Test @ vol=2) strategies, we train the 2-Action market maker to decide between providing bilateral quotes or no quotes. \textbf{4-Action MM (Train @ vol=2)}: Under volatility=2, using the adversary (Train \& Test @ vol=2) and Always Quoting MM (Train \& Test @ vol=2) strategies, we train the 4-Action market maker to decide between providing bilateral quotes, unilateral quotes, or no quotes. \textbf{4-Action MM (Train @ vol=200)}: Under volatility=200, using the adversary (Train \& Test @ vol=2) and Always Quoting MM (Train \& Test @ vol=2) strategies, we train the 4-Action market maker to decide between providing bilateral quotes, unilateral quotes, or no quotes.

Following this, we evaluate the performance of each agent in their respective testing environments based on the evaluation metrics outlined in Section \ref{subsec:metrics}. Upon completing the above process, we change the type of adversary or adjust the risk coefficients and repeat the procedure.

\vspace{-0.1em}

\section{Evaluation}

\subsection{Metrics}
\label{subsec:metrics}
Since the reward function of all agents is influenced by risk parameters, this paper evaluates the performance of all market-making agents under seven different sets of risk coefficients, as shown in Tables \ref{RN} - \ref{zeta0.001}.

For each set of risk coefficients, each type of market maker has strategies generated through reinforcement learning in various adversarial environments for training. All strategies were then evaluated 100 times, each with 1000 episodes, in a fixed adversarial environment for testing. The tables present statistical results of the market makers' performance across multiple dimensions.

Specifically, each table includes the performance of six market makers: Always Quoting MM (Train \& Test @ vol=2), 2-Action MM (Train \& Test @ vol=2), 4-Action MM (Train \& Test @ vol=2), 4-Action MM (Train @ vol=2, Test @ vol=200), and 4-Action MM (Train \& Test @ vol=200). These six market makers differ significantly in their action spaces and in the volatility ($\sigma$) values of their training and testing environments. For example, the 4-Action MM (Train \& Test @ vol=2) refers to a 4-action market maker trained and tested in an environment with Volatility=2, using adversarial agents and Always Quoting MM strategies also trained in an environment with Volatility=2. The 4-Action MM (Train @ vol=2, Test @ vol=200) follows the same training process but is tested in an environment with Volatility=200. Another distinct group is the 4-Action MM (Train \& Test @ vol=200), which is trained in an environment with Volatility=200, using adversarial agents and Always Quoting MM strategies trained in an environment with Volatility=2, but ultimately tested in an environment with Volatility=2.

The performance of the market makers is primarily measured by four metrics: Terminal Wealth, Sharpe Ratio, Terminal Inventory, and Quoting Ratio. Terminal Wealth refers to the agent’s distribution of episodic terminal wealth, presented as ‘the mean ± the variance’, i.e., $E(\prod_N)$ ± $\sigma(\prod_N)$. The Sharpe Ratio, defined as $\frac{E(\prod_N)}{\sigma(\prod_N)}$, measures the reward per unit of risk. A high Sharpe Ratio, indicating a large mean and low variance of wealth, is a key indicator of good market maker performance. Terminal Inventory refers to the distribution of terminal inventory, $H_N$, providing insight into the agent's inventory management. Additionally, for 2-Action MM and 4-Action MM, the Quoting Ratio is an important performance metric. The quoting ratio in the tables is presented as 'no-quote ratio + bilateral quote ratio + ask-only ratio + bid-only ratio'.

\begin{table}[h]
\centering
\caption{Market makers trained with $\eta = 0.0$ and $\zeta = 0.0$ (RN)}
\label{RN}
\resizebox{\columnwidth}{!}
{%
\begin{tabular}{|c|c|c|c|c|c|c|}
\hline
Adversary &
  Desiderata &
  \begin{tabular}[c]{@{}c@{}}Always Quoting   MM\\      (Train @ vol=2,\\      Test @ vol=2)\end{tabular} &
  \begin{tabular}[c]{@{}c@{}}2-Action   MM\\      (Train @ vol=2,\\      Test @ vol=2)\end{tabular} &
  \begin{tabular}[c]{@{}c@{}}4-Action   MM\\      (Train @ vol=2,\\      Test @ vol=2)\end{tabular} &
  \begin{tabular}[c]{@{}c@{}}4-Action   MM\\      (Train @ vol=2,\\      Test @ vol=200)\end{tabular} &
  \begin{tabular}[c]{@{}c@{}}4-Action   MM\\      (Train @ vol=200, \\      Test @ vol=200)\end{tabular} \\ \hline
\multirow{4}{*}{Fix}    & Term.   Wealth  & 2.1945±3.2613  & 2.2619±3.2623         & 2.2389±3.2875         & 2.2501±3.2909        & 2.0170±3.2892         \\ \cline{2-7} 
                        & Sharpe          & 0.6729         & 0.6934                & 0.6810                & 0.6837               & 0.6132                \\ \cline{2-7} 
                        & Term. Inventory & 1.4109±0.9533  & 1.4217±0.9506         & 1.4271±0.9396         & 1.4276±0.9419        & 1.5679±1.0039         \\ \cline{2-7} 
                        & Quoting Ratio   & -               & 0.00+100.00+0.00+0.00 & 0.00+99.17+0.00+0.83  & 0.00+99.29+0.00+0.71 & 0.00+74.19+0.00+25.81 \\ \hline
\multirow{4}{*}{Random} & Term.   Wealth  & 2.9423±3.7090  & 2.9770±3.5750         & 2.9849±3.6404         & 2.9272±3.7251        & 2.4422±3.9800         \\ \cline{2-7} 
                        & Sharpe          & 0.7933         & 0.8327                & 0.8200                & 0.7858               & 0.6136                \\ \cline{2-7} 
                        & Term. Inventory & -0.7374±1.9390 & -0.7360±2.0294        & -0.5960±1.8485        & -0.6977±1.9465       & -1.3190±2.4119        \\ \cline{2-7} 
                        & Quoting Ratio   & -              & 0.00+100.00+0.00+0.00 & 0.00+100.00+0.00+0.00 & 0.06+99.94+0.00+0.00 & 0.00+73.61+26.39+0.00 \\ \hline
\multirow{4}{*}{A}      & Term.   Wealth  & 2.0423±2.5369  & 2.0315±2.4914         & 2.0128±2.4954         & 2.0392±2.5512        & 1.6238±2.3433         \\ \cline{2-7} 
                        & Sharpe          & 0.8050         & 0.8154                & 0.8066                & 0.7993               & 0.6930                \\ \cline{2-7} 
                        & Term. Inventory & -0.1345±0.8514 & -0.1300±0.8445        & -0.1315±0.8425        & -0.1204±0.8605       & -0.0051±0.9256        \\ \cline{2-7} 
                        & Quoting Ratio   & -              & 0.01+99.99+0.00+0.00  & 0.00+99.88+0.12+0.00  & 0.00+99.90+0.10+0.00 & 20.85+77.27+1.88+0.00 \\ \hline
\multirow{4}{*}{B}      & Term.   Wealth  & 2.2895±2.7648  & 2.3614±2.7903         & 2.1328±2.6252         & 2.3110±2.7580        & 1.8890±2.4494         \\ \cline{2-7} 
                        & Sharpe          & 0.8281         & 0.8463                & 0.8124                & 0.8379               & 0.7712                \\ \cline{2-7} 
                        & Term. Inventory & 0.7640±0.8546  & 0.7570±0.8698         & 0.7520±0.8675         & 0.7481±0.8689        & 0.2670±1.1856         \\ \cline{2-7} 
                        & Quoting Ratio   & -              & 0.00+100.00+0.00+0.00 & 1.01+98.70+0.00+0.29  & 0.82+98.95+0.00+0.23 & 0.00+69.59+30.41+0.00 \\ \hline
\multirow{4}{*}{K}      & Term.   Wealth  & 2.9030±3.7789  & 2.9974±3.8130         & 3.0503±3.8610         & 2.8917±3.7534        & 2.4803±3.5783         \\ \cline{2-7} 
                        & Sharpe          & 0.7682         & 0.7861                & 0.7900                & 0.7704               & 0.6932                \\ \cline{2-7} 
                        & Term. Inventory & 0.0200±2.0204  & 0.0106±2.0316         & 0.0030±2.0349         & -0.0250±2.0407       & -0.1610±2.1239        \\ \cline{2-7} 
                        & Quoting Ratio   & -              & 0.00+100.00+0.00+0.00 & 0.00+97.65+2.35+0.00  & 0.00+97.26+2.74+0.00 & 0.00+62.80+37.20+0.00 \\ \hline
\multirow{4}{*}{All}    & Term.   Wealth  & 3.8587±3.7280  & 3.8373±3.5741         & 3.8669±3.6035         & 3.1140±2.9786        & 3.4088±3.7417         \\ \cline{2-7} 
                        & Sharpe          & 1.0351         & 1.0736                & 1.0731                & 1.0455               & 0.9110                \\ \cline{2-7} 
                        & Term. Inventory & 1.3965±1.7463  & 1.3982±1.7223         & 0.1080±1.0846         & -0.5299±0.6196       & 1.8780±1.6683         \\ \cline{2-7} 
                        & Quoting Ratio   & -              & 0.00+100.00+0.00+0.00 & 0.23+99.77+0.00+0.00  & 0.63+99.37+0.00+0.00 & 0.00+78.20+0.00+21.80 \\ \hline
\end{tabular}%
}
\end{table}

\begin{table}[h]
\centering
\caption{Market makers trained with $\eta = 0.01$ and $\zeta = 0.0$}
\label{eta0.01}
\resizebox{\columnwidth}{!}{%
\begin{tabular}{|c|c|c|c|c|c|c|}
\hline
Adversary &
  Desiderata &
  \begin{tabular}[c]{@{}c@{}}Always Quoting   MM\\      (Train @ vol=2,\\      Test @ vol=2)\end{tabular} &
  \begin{tabular}[c]{@{}c@{}}2-Action   MM\\      (Train @ vol=2,\\      Test @ vol=2)\end{tabular} &
  \begin{tabular}[c]{@{}c@{}}4-Action   MM\\      (Train @ vol=2,\\      Test @ vol=2)\end{tabular} &
  \begin{tabular}[c]{@{}c@{}}4-Action   MM\\      (Train @ vol=2,\\      Test @ vol=200)\end{tabular} &
  \begin{tabular}[c]{@{}c@{}}4-Action   MM\\      (Train @ vol=200, \\      Test @ vol=200)\end{tabular} \\ \hline
\multirow{4}{*}{Fix}    & Term.   Wealth  & 2.1288±2.8318  & 2.0743±2.7506         & 2.1134±2.8823         & 2.0888±2.7243         & 1.9595±2.6252         \\ \cline{2-7} 
                        & Sharpe          & 0.7518         & 0.7541                & 0.7332                & 0.7667                & 0.7464                \\ \cline{2-7} 
                        & Term. Inventory & -0.0599±1.3094 & -0.0571±1.2801        & -0.0640±1.2775        & -0.0733±1.2869        & -0.0830±1.1714        \\ \cline{2-7} 
                        & Quoting Ratio   & -              & 0.00+100.00+0.00+0.00 & 0.00+99.99+0.01+0.00  & 0.00+99.93+0.07+0.00  & 0.76+69.74+29.37+0.13 \\ \hline
\multirow{4}{*}{Random} & Term.   Wealth  & 2.9421±2.8311  & 2.9964±2.7446         & 2.4722±2.4654         & 2.4476±2.4654         & 2.6421±2.6762         \\ \cline{2-7} 
                        & Sharpe          & 1.0392         & 1.0917                & 1.0028                & 0.9928                & 0.9872                \\ \cline{2-7} 
                        & Term. Inventory & 0.1847±1.0268  & 0.2130±1.0496         & -0.6050±0.8337        & -0.2850±0.6350        & 0.0640±1.0908         \\ \cline{2-7} 
                        & Quoting Ratio   & -              & 0.78+99.22+0.00+0.00  & 0.28+99.72+0.00+0.00  & 3.98+96.02+0.00+0.00  & 0.00+86.80+13.20+0.00 \\ \hline
\multirow{4}{*}{All}    & Term.   Wealth  & 3.4945±2.9118  & 3.4886±2.9243         & 3.5789±3.0588         & 3.5145±2.9112         & 2.8375±2.6318         \\ \cline{2-7} 
                        & Sharpe          & 1.2001         & 1.1930                & 1.1700                & 1.2072                & 1.0782                \\ \cline{2-7} 
                        & Term. Inventory & -0.1795±1.1329 & -0.1586±1.1299        & -0.1570±1.1333        & -0.1546±1.1318        & -0.2530±0.9834        \\ \cline{2-7} 
                        & Quoting Ratio   & -              & 0.00+100.00+0.00+0.00 & 0.00+100.00+0.00+0.00 & 0.00+100.00+0.00+0.00 & 0.00+81.41+18.59+0.00 \\ \hline
\end{tabular}%
}
\end{table}

\begin{table}[h]
\centering
\caption{Market makers trained with $\eta = 0.1$ and $\zeta = 0.0$}
\label{eta0.1}
\resizebox{\columnwidth}{!}{%
\begin{tabular}{|c|c|c|c|c|c|c|}
\hline
Adversary &
  Desiderata &
  \begin{tabular}[c]{@{}c@{}}Always Quoting   MM\\      (Train @ vol=2,\\      Test @ vol=2)\end{tabular} &
  \begin{tabular}[c]{@{}c@{}}2-Action   MM\\      (Train @ vol=2,\\      Test @ vol=2)\end{tabular} &
  \begin{tabular}[c]{@{}c@{}}4-Action   MM\\      (Train @ vol=2,\\      Test @ vol=2)\end{tabular} &
  \begin{tabular}[c]{@{}c@{}}4-Action   MM\\      (Train @ vol=2,\\      Test @ vol=200)\end{tabular} &
  \begin{tabular}[c]{@{}c@{}}4-Action   MM\\      (Train @ vol=200, \\      Test @ vol=200)\end{tabular} \\ \hline
\multirow{4}{*}{Fix}    & Term.   Wealth  & 2.8758±3.9043  & 2.8903±3.9577         & 2.7903±4.1780         & 2.8136±3.9301         & 2.8088±3.6980         \\ \cline{2-7} 
                        & Sharpe          & 0.7366         & 0.7303                & 0.6679                & 0.7159                & 0.7596                \\ \cline{2-7} 
                        & Term. Inventory & -0.0468±2.4101 & -0.0695±2.3688        & 0.0490±2.4455         & -0.1286±2.3282        & -0.0130±2.3188        \\ \cline{2-7} 
                        & Quoting Ratio   & -              & 0.00+100.00+0.00+0.00 & 0.00+98.60+1.32+0.09  & 0.00+98.25+1.68+0.07  & 0.00+100.00+0.00+0.00 \\ \hline
\multirow{4}{*}{Random} & Term.   Wealth  & 3.5162±3.9344  & 3.3342±3.7026         & 2.3575±2.6479         & 3.2742±2.6440         & 2.8890±3.7461         \\ \cline{2-7} 
                        & Sharpe          & 0.8937         & 0.9005                & 0.8903                & 1.2384                & 0.7712                \\ \cline{2-7} 
                        & Term. Inventory & 0.2517±1.2195  & 0.1950±1.1211         & 0.0610±0.8951         & 0.0722±0.8794         & 0.2670±1.1856         \\ \cline{2-7} 
                        & Quoting Ratio   & -              & 0.39+99.61+0.00+0.00  & 0.00+98.83+0.65+0.52  & 0.00+92.98+6.54+0.48  & 0.00+69.59+30.41+0.00 \\ \hline
\multirow{4}{*}{All}    & Term.   Wealth  & 2.6306±2.8617  & 2.5936±2.7803         & 2.7435±2.9880         & 2.6100±2.8343         & 2.2474±2.7312         \\ \cline{2-7} 
                        & Sharpe          & 0.9192         & 0.9329                & 0.9182                & 0.9209                & 0.8228                \\ \cline{2-7} 
                        & Term. Inventory & -0.2475±1.2561 & -0.2459±1.2465        & -0.3150±1.3273        & -0.2673±1.2590        & -0.4890±1.4683        \\ \cline{2-7} 
                        & Quoting Ratio   & -              & 0.00+100.00+0.00+0.00 & 0.00+100.00+0.00+0.00 & 0.00+100.00+0.00+0.00 & 0.00+86.19+13.81+0.00 \\ \hline
\end{tabular}%
}
\end{table}

\begin{table}[h]
\centering
\caption{Market makers trained with $\eta = 0.5$ and $\zeta   = 0.0$}
\label{eta0.5}
\resizebox{\columnwidth}{!}{%
\begin{tabular}{|c|c|c|c|c|c|c|}
\hline
Adversary &
  Desiderata &
  \begin{tabular}[c]{@{}c@{}}Always Quoting   MM\\      (Train @ vol=2,\\      Test @ vol=2)\end{tabular} &
  \begin{tabular}[c]{@{}c@{}}2-Action   MM\\      (Train @ vol=2,\\      Test @ vol=2)\end{tabular} &
  \begin{tabular}[c]{@{}c@{}}4-Action   MM\\      (Train @ vol=2,\\      Test @ vol=2)\end{tabular} &
  \begin{tabular}[c]{@{}c@{}}4-Action   MM\\      (Train @ vol=2,\\      Test @ vol=200)\end{tabular} &
  \begin{tabular}[c]{@{}c@{}}4-Action   MM\\      (Train @ vol=200, \\      Test @ vol=200)\end{tabular} \\ \hline
\multirow{4}{*}{Fix}    & Term.   Wealth  & 2.0725±2.6569  & 2.0763±2.6441         & 2.0298±2.5432        & 1.9588±2.6125        & 1.7946±2.5060         \\ \cline{2-7} 
                        & Sharpe          & 0.7800         & 0.7853                & 0.7981               & 0.7498               & 0.7162                \\ \cline{2-7} 
                        & Term. Inventory & 0.5562±0.8874  & 0.5460±0.8879         & 0.4710±0.9577        & 0.5017±0.9972        & 0.4940±0.7429         \\ \cline{2-7} 
                        & Quoting Ratio   & -              & 0.00+100.00+0.00+0.00 & 0.00+95.60+3.33+1.07 & 0.00+95.40+3.55+1.05 & 3.07+43.76+53.17+0.00 \\ \hline
\multirow{4}{*}{Random} & Term.   Wealth  & 3.3054±3.8613  & 3.3755±3.8794         & 2.3934±2.7395        & 2.1739±2.4552        & 1.9857±2.6252         \\ \cline{2-7} 
                        & Sharpe          & 0.8560         & 0.8701                & 0.8737               & 0.8854               & 0.7564                \\ \cline{2-7} 
                        & Term. Inventory & -1.2291±1.9555 & -1.1750±1.9592        & -0.1140±1.0281       & 0.4760±0.8255        & -0.0830±1.1714        \\ \cline{2-7} 
                        & Quoting Ratio   & -              & 0.00+100.00+0.00+0.00 & 0.04+99.96+0.00+0.00 & 0.00+98.27+0.00+1.73 & 0.76+73.74+25.37+0.13 \\ \hline
\multirow{4}{*}{All}    & Term.   Wealth  & 2.3682±2.5283  & 2.3510±2.5355         & 2.3241±2.4664        & 2.3175±2.4661        & 1.8467±2.2641         \\ \cline{2-7} 
                        & Sharpe          & 0.9367         & 0.9272                & 0.9423               & 0.9397               & 0.8157                \\ \cline{2-7} 
                        & Term. Inventory & 0.5521±0.7390  & 0.5603±0.7311         & 0.5420±0.7143        & 0.5556±0.7282        & 0.8850±1.0079         \\ \cline{2-7} 
                        & Quoting Ratio   & -              & 0.00+100.00+0.00+0.00 & 0.00+92.20+0.39+7.41 & 0.00+92.51+0.38+7.10 & 0.00+92.59+5.85+1.56  \\ \hline
\end{tabular}%
}
\end{table}

\begin{table}[h]
\centering
\caption{Market makers trained with $\eta = 1.0$ and $\zeta = 0.0$}
\label{eta1}
\resizebox{\columnwidth}{!}{%
\begin{tabular}{|c|c|c|c|c|c|c|}
\hline
Adversary &
  Desiderata &
  \begin{tabular}[c]{@{}c@{}}Always Quoting   MM\\      (Train @ vol=2,\\      Test @ vol=2)\end{tabular} &
  \begin{tabular}[c]{@{}c@{}}2-Action   MM\\      (Train @ vol=2,\\      Test @ vol=2)\end{tabular} &
  \begin{tabular}[c]{@{}c@{}}4-Action   MM\\      (Train @ vol=2,\\      Test @ vol=2)\end{tabular} &
  \begin{tabular}[c]{@{}c@{}}4-Action   MM\\      (Train @ vol=2,\\      Test @ vol=200)\end{tabular} &
  \begin{tabular}[c]{@{}c@{}}4-Action   MM\\      (Train @ vol=200, \\      Test @ vol=200)\end{tabular} \\ \hline
\multirow{4}{*}{Fix}    & Term.   Wealth  & 2.3124±2.9290  & 2.3230±2.9271         & 2.4013±2.9184        & 2.3635±2.9494        & 1.2129±1.6319         \\ \cline{2-7} 
                        & Sharpe          & 0.7895         & 0.7936                & 0.8228               & 0.8013               & 0.7433                \\ \cline{2-7} 
                        & Term. Inventory & 0.3423±1.2215  & 0.3434±1.2429         & 0.2910±1.1993        & 0.3307±1.2302        & 0.1100±0.8921         \\ \cline{2-7} 
                        & Quoting Ratio   & -              & 0.00+100.00+0.00+0.00 & 0.02+99.98+0.00+0.00 & 0.03+99.97+0.00+0.00 & 45.45+46.96+7.30+0.29 \\ \hline
\multirow{4}{*}{Random} & Term.   Wealth  & 2.8369±2.8352  & 2.7148±2.6612         & 2.7694±2.6248        & 3.1713±2.7002        & 2.2828±2.1431         \\ \cline{2-7} 
                        & Sharpe          & 1.0006         & 1.0201                & 1.0551               & 1.1745               & 1.0652                \\ \cline{2-7} 
                        & Term. Inventory & -0.2089±0.9267 & -0.1970±0.8833        & -0.1020±0.8097       & -0.1767±0.8361       & 0.1960±0.6750         \\ \cline{2-7} 
                        & Quoting Ratio   & -              & 2.55+97.45+0.00+0.00  & 0.00+94.94+0.00+5.07 & 0.00+94.63+0.00+5.37 & 0.00+92.83+6.65+0.51  \\ \hline
\multirow{4}{*}{All}    & Term.   Wealth  & 2.6801±2.5553  & 2.7281±2.5363         & 2.8309±2.5555        & 2.6384±2.5127        & 2.6385±2.4315         \\ \cline{2-7} 
                        & Sharpe          & 1.0489         & 1.0756                & 1.1077               & 1.0500               & 1.0851                \\ \cline{2-7} 
                        & Term. Inventory & -0.5668±0.9262 & -0.5818±0.9181        & -0.5540±0.9269       & -0.6012±0.8993       & -0.6010±0.8762        \\ \cline{2-7} 
                        & Quoting Ratio   & -              & 0.00+100.00+0.00+0.00 & 0.02+91.76+3.85+4.38 & 0.02+92.02+3.73+4.23 & 0.00+100.00+0.00+0.00 \\ \hline
\end{tabular}%
}
\end{table}

\begin{table}[h]
\centering
\caption{Market makers trained with $\eta = 0.0$ and $\zeta = 0.01$}
\label{zeta0.01}
\resizebox{\columnwidth}{!}{%
\begin{tabular}{|c|c|c|c|c|c|c|}
\hline
Adversary &
  Desiderata &
  \begin{tabular}[c]{@{}c@{}}Always Quoting   MM\\      (Train @ vol=2,\\      Test @ vol=2)\end{tabular} &
  \begin{tabular}[c]{@{}c@{}}2-Action   MM\\      (Train @ vol=2,\\      Test @ vol=2)\end{tabular} &
  \begin{tabular}[c]{@{}c@{}}4-Action   MM\\      (Train @ vol=2,\\      Test @ vol=2)\end{tabular} &
  \begin{tabular}[c]{@{}c@{}}4-Action   MM\\      (Train @ vol=2,\\      Test @ vol=200)\end{tabular} &
  \begin{tabular}[c]{@{}c@{}}4-Action   MM\\      (Train @ vol=200, \\      Test @ vol=200)\end{tabular} \\ \hline
\multirow{4}{*}{Fix}    & Term.   Wealth  & 2.3571±2.7381  & 2.4111±2.7926         & 2.3755±2.6941         & 2.3536±2.7629         & 1.8631±2.3480         \\ \cline{2-7} 
                        & Sharpe          & 0.8608         & 0.8634                & 0.8817                & 0.8519                & 0.7935                \\ \cline{2-7} 
                        & Term. Inventory & 0.3905±0.9272  & 0.4020±0.9404         & 0.3840±0.9124         & 0.4061±0.9240         & -0.2970±0.5694        \\ \cline{2-7} 
                        & Quoting Ratio   & -              & 0.07+99.93+0.00+0.00  & 0.00+99.17+0.83+0.00  & 0.00+99.29+0.71+0.00  & 1.14+31.44+67.40+0.01 \\ \hline
\multirow{4}{*}{Random} & Term.   Wealth  & 2.5684±2.3927  & 2.5133±2.2925         & 2.4560±2.2804         & 2.4538±2.3210         & 2.4040±2.3716         \\ \cline{2-7} 
                        & Sharpe          & 1.0734         & 1.0963                & 1.0770                & 1.0572                & 1.0137                \\ \cline{2-7} 
                        & Term. Inventory & -0.2037±0.6080 & -0.2070±0.6034        & -0.2090±0.6351        & -0.2050±0.6200        & -0.3590±0.8162        \\ \cline{2-7} 
                        & Quoting Ratio   & -              & 0.25+99.75+0.00+0.00  & 1.56+98.44+0.00+0.00  & 1.34+98.66+0.00+0.00  & 0.20+83.01+16.79+0.00 \\ \hline
\multirow{4}{*}{All}    & Term.   Wealth  & 2.3433±2.0926  & 2.3178±2.1019         & 2.2747±2.0776         & 2.3477±2.0786         & 2.2071±2.0902         \\ \cline{2-7} 
                        & Sharpe          & 1.1198         & 1.1027                & 1.0949                & 1.1295                & 1.0559                \\ \cline{2-7} 
                        & Term. Inventory & 0.1312±0.6071  & 0.1231±0.6001         & 0.1130±0.6182         & 0.1333±0.5976         & 0.1460±0.6393         \\ \cline{2-7} 
                        & Quoting Ratio   & -              & 0.00+100.00+0.00+0.00 & 0.00+100.00+0.00+0.00 & 0.00+100.00+0.00+0.00 & 0.00+99.52+0.46+0.02  \\ \hline
\end{tabular}%
}
\end{table}

\begin{table}[h]
\centering
\caption{Market makers trained with $\eta = 0.0$ and $\zeta = 0.001$}
\label{zeta0.001}
\resizebox{\columnwidth}{!}{%
\begin{tabular}{|c|c|c|c|c|c|c|}
\hline
Adversary &
  Desiderata &
  \begin{tabular}[c]{@{}c@{}}Always Quoting   MM\\      (Train @ vol=2,\\      Test @ vol=2)\end{tabular} &
  \begin{tabular}[c]{@{}c@{}}2-Action   MM\\      (Train @ vol=2,\\      Test @ vol=2)\end{tabular} &
  \begin{tabular}[c]{@{}c@{}}4-Action   MM\\      (Train @ vol=2,\\      Test @ vol=2)\end{tabular} &
  \begin{tabular}[c]{@{}c@{}}4-Action   MM\\      (Train @ vol=2,\\      Test @ vol=200)\end{tabular} &
  \begin{tabular}[c]{@{}c@{}}4-Action   MM\\      (Train @ vol=200, \\      Test @ vol=200)\end{tabular} \\ \hline
\multirow{4}{*}{Fix}    & Term.   Wealth  & 2.8302±3.0068  & 2.7519±2.6668         & 2.7102±2.7434         & 2.8019±2.6934         & 1.9953±2.4853         \\ \cline{2-7} 
                        & Sharpe          & 0.9413         & 1.0319                & 0.9879                & 1.0403                & 0.8029                \\ \cline{2-7} 
                        & Term. Inventory & -0.7793±0.8051 & -0.7630±0.7961        & -0.7710±0.7580        & -0.7483±0.8185        & -0.8620±0.6950        \\ \cline{2-7} 
                        & Quoting Ratio   & -              & 0.00+100.00+0.00+0.00 & 0.00+100.00+0.00+0.00 & 0.00+100.00+0.00+0.00 & 0.92+48.77+50.31+0.00 \\ \hline
\multirow{4}{*}{Random} & Term.   Wealth  & 2.3603±2.3478  & 2.4809±2.3535         & 2.3478±2.2933         & 2.3341±2.2821         & 2.3484±2.3196         \\ \cline{2-7} 
                        & Sharpe          & 1.0053         & 1.0541                & 1.0237                & 1.0228                & 1.0124                \\ \cline{2-7} 
                        & Term. Inventory & 0.3531±0.7931  & 0.3430±0.8008         & 0.3060±0.8273         & 0.3116±0.8154         & 0.3150±0.7898         \\ \cline{2-7} 
                        & Quoting Ratio   & -              & 0.00+100.00+0.00+0.00 & 1.05+98.92+0.03+0.00  & 1.08+98.91+0.01+0.00  & 0.10+99.84+0.06+0.00  \\ \hline
\multirow{4}{*}{All}    & Term.   Wealth  & 3.4308±2.6641  & 3.4338±2.6881         & 3.2481±2.7331         & 3.4441±2.6385         & 1.8499±1.7792         \\ \cline{2-7} 
                        & Sharpe          & 1.2878         & 1.2774                & 1.1885                & 1.3053                & 1.0397                \\ \cline{2-7} 
                        & Term. Inventory & -0.2228±0.9603 & -0.2008±0.9654        & -0.2350±0.9443        & -0.2167±0.9681        & 0.3410±0.5067         \\ \cline{2-7} 
                        & Quoting Ratio   & -              & 0.00+100.00+0.00+0.00 & 0.00+99.94+0.06+0.00  & 0.00+99.98+0.02+0.00  & 0.00+74.64+0.00+25.36 \\ \hline
\end{tabular}%
}
\end{table}

\subsection{Results Analysis}

\begin{figure}
        \centering
        \includegraphics[width=\columnwidth]{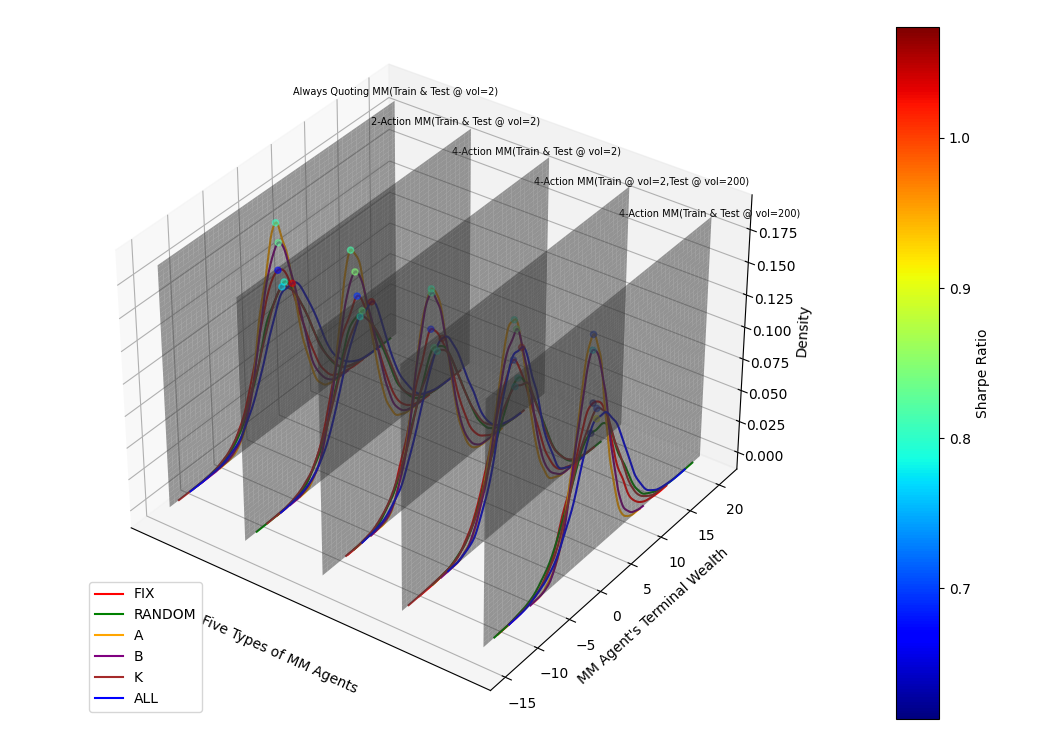}
        \caption{\small Performance of Five Types of MM Agents (\small $\eta=0.0, \zeta=0.0$)} 
        \label{figure_ALL}
    \end{figure}   

\begin{figure}
        \centering
        \includegraphics[width=\columnwidth]{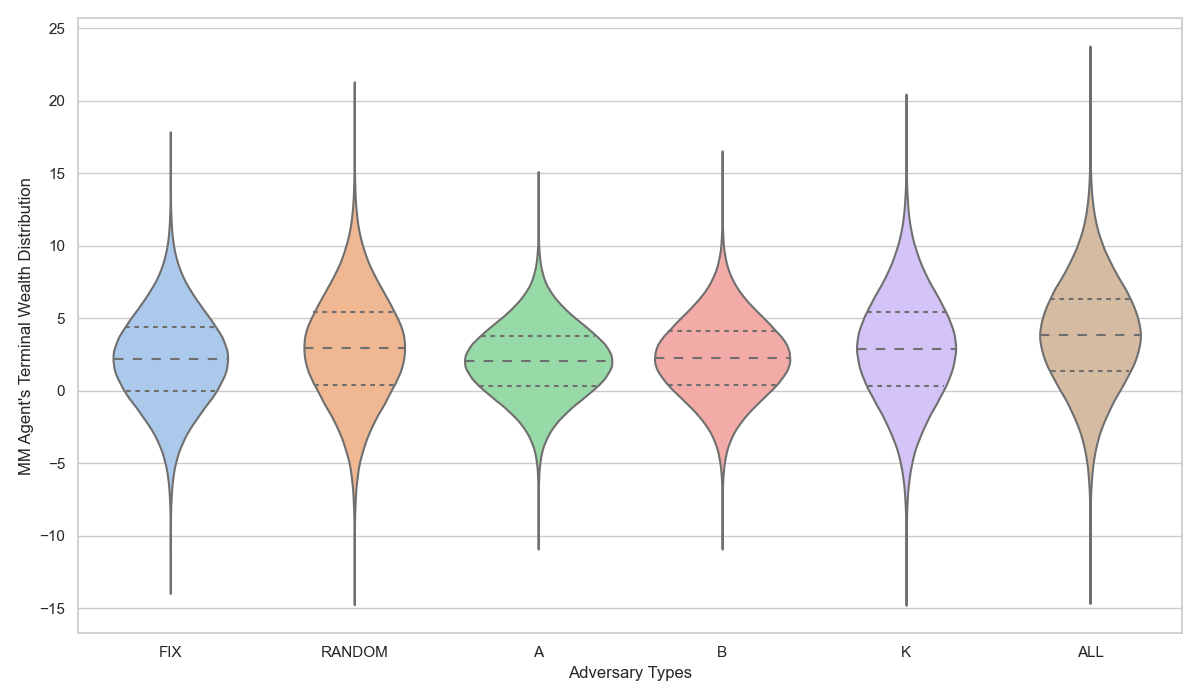}
        \caption{\small Performance of Always Quoting MM  (\small $\eta=0.0, \zeta=0.0$)}
        \label{Always Quoting MM}
    \end{figure}

\begin{figure}
        \centering
        \includegraphics[width=\columnwidth]{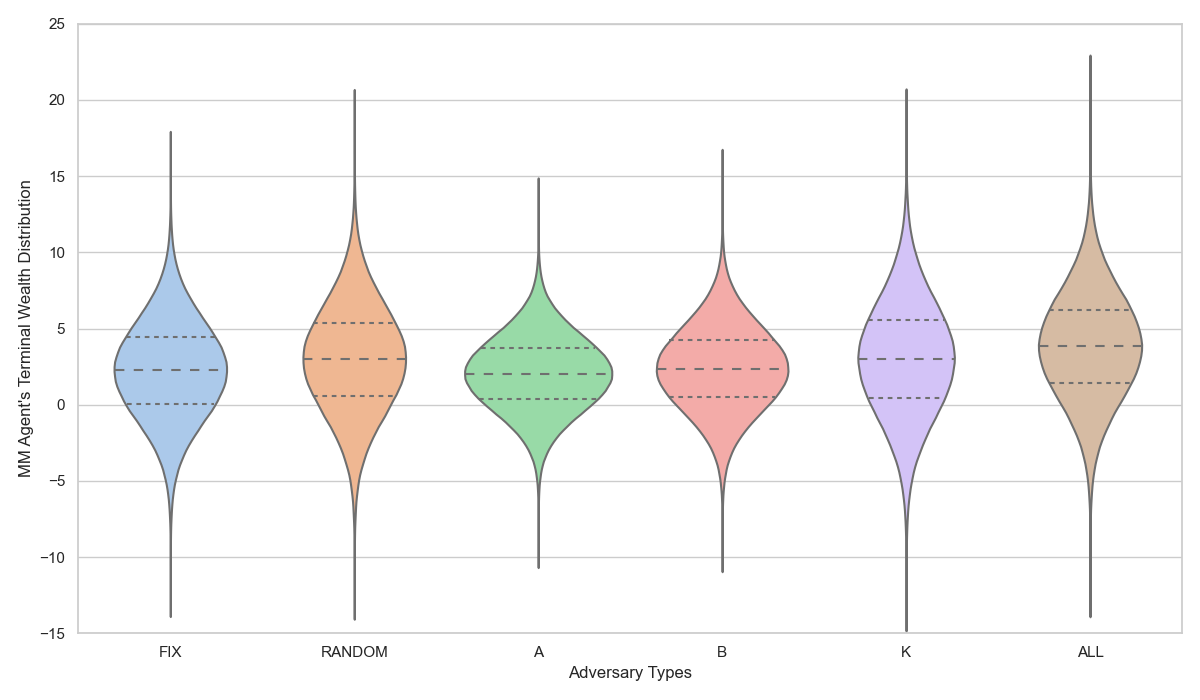}
        \caption{\small Performance of 2-Action MM (Train \& Test @ vol=2)  (\small $\eta=0.0, \zeta=0.0$)}
        \label{2action}
    \end{figure}

\begin{figure}
        \centering
        \includegraphics[width=\columnwidth]{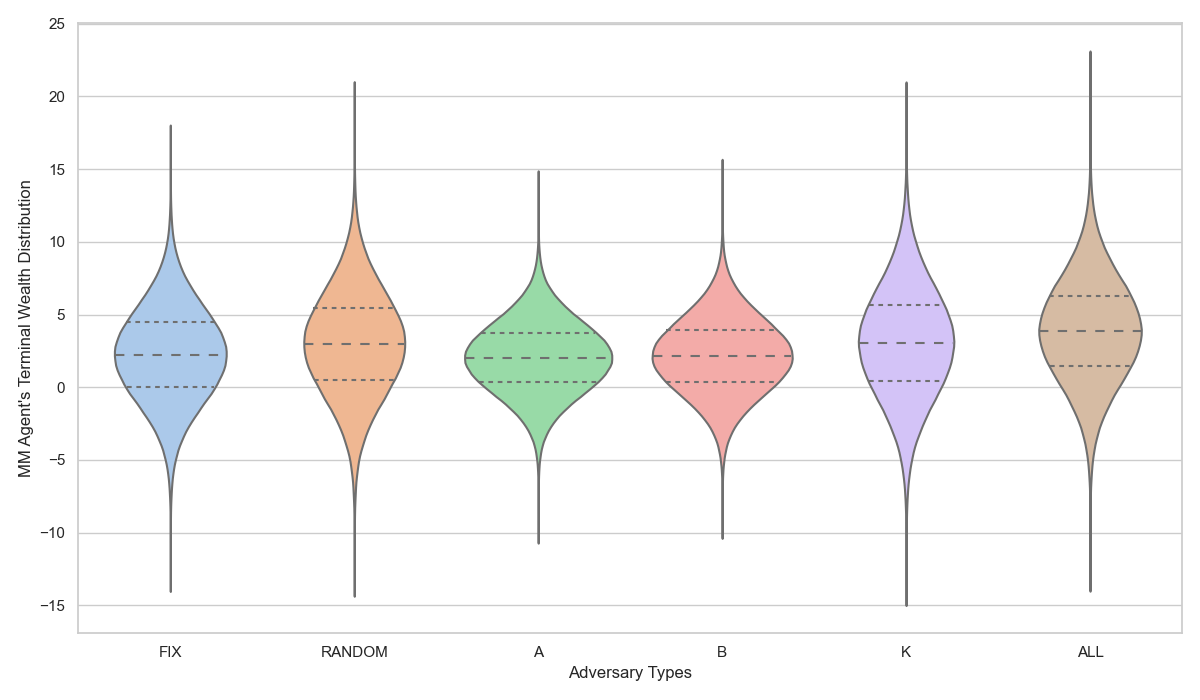}
        \caption{\small Performance of 4-Action MM (Train \& Test @ vol=2)  (\small $\eta=0.0, \zeta=0.0$)}
        \label{4Action_1}
    \end{figure}

\begin{figure}
        \centering
        \includegraphics[width=\columnwidth]{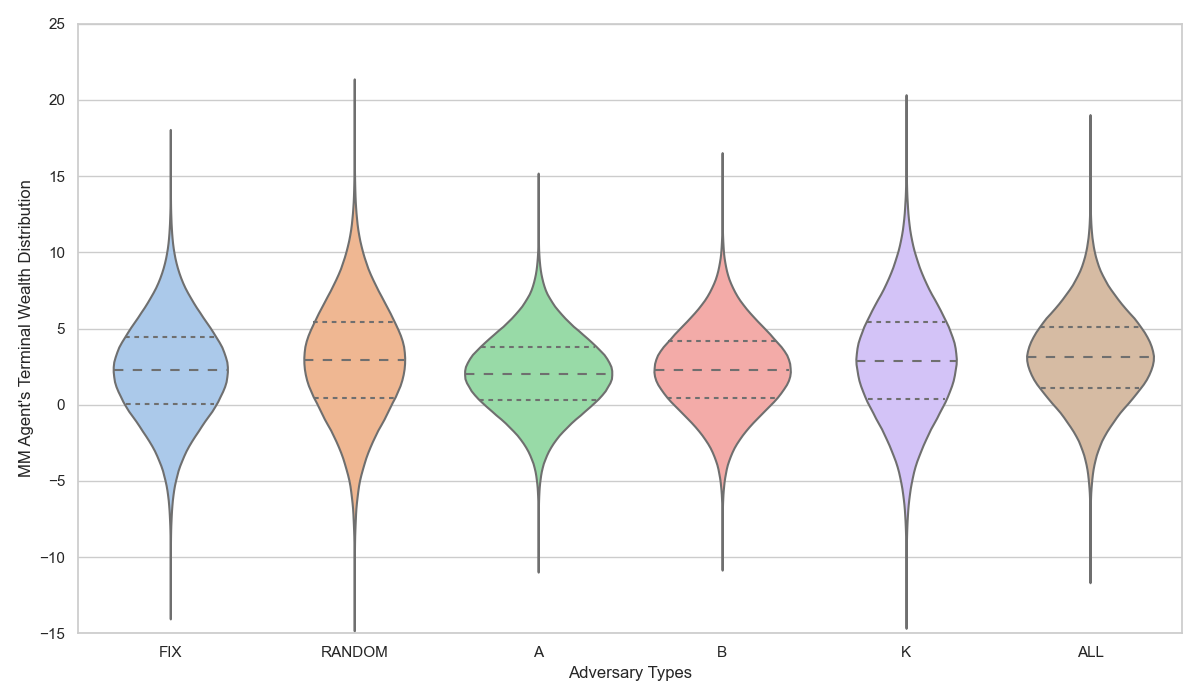}
        \caption{\small Performance of 4-Action MM (Train @ vol=2, Test @ vol=200)  (\small $\eta=0.0, \zeta=0.0$)}
        \label{4Action_2}
    \end{figure}

\begin{figure}
        \centering
        \includegraphics[width=\columnwidth]{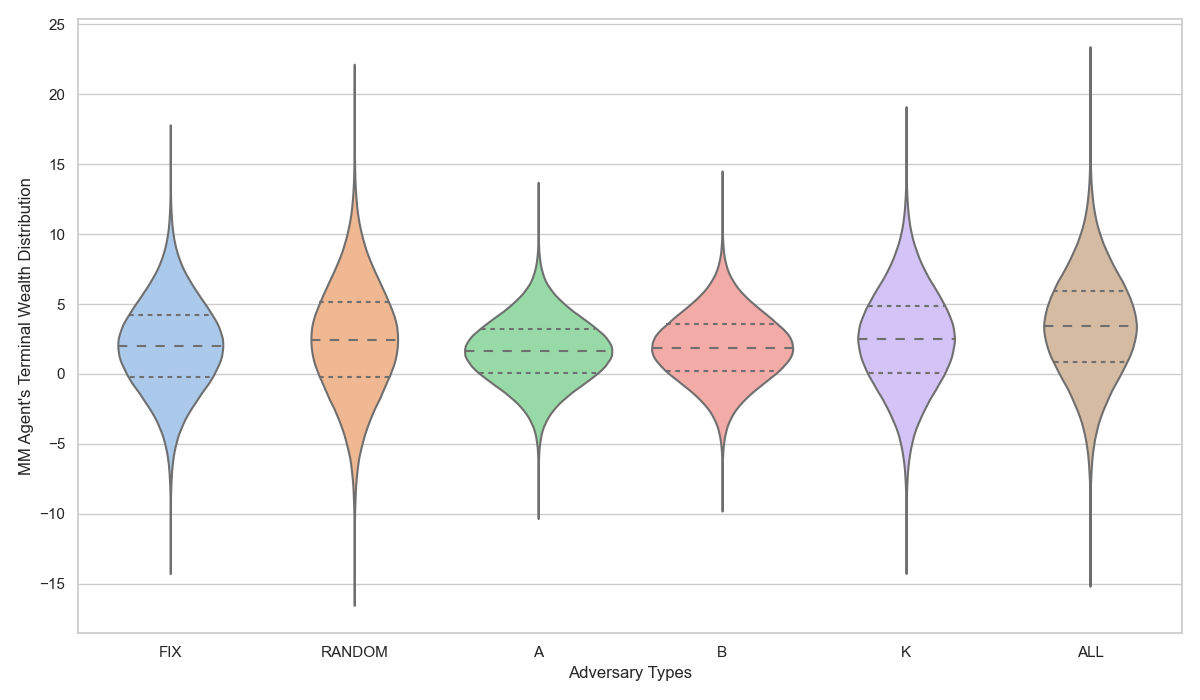}
        \caption{\small Performance of 4-Action MM (Train \& Test @ vol=200)  (\small $\eta=0.0, \zeta=0.0$)}
        \label{4Action_3}
    \end{figure}

\noindent\textbf{Always Quoting MM (Train \& Test @ vol=2).} In this paper, the "Always Quoting" market maker (MM) strategy differs significantly from the models in \cite{spooner2020robust} and \cite{wang2023robust} in its order arrival process. While those studies use the Poisson process, this paper employs the Hawkes process.

In terms of performance, the "Always Quoting" MM presented here exhibits similar trends to those found in \cite{spooner2020robust} and \cite{wang2023robust}: under All Adversary training with three controlled parameters, it achieves the best Sharpe ratio, outperforming the results from Random Adversary training, which in turn are superior to those from Fixed Adversary training. Additionally, the Sharpe ratio for the MM trained with B Adversary is generally higher than that for Random Adversary training. However, no significant advantage is observed for the MM under A Adversary and K Adversary adversarial training conditions.

Despite these observations, the market maker in this study performs worse overall in terms of wealth mean $E(\prod_N)$ and Sharpe ratio compared to results using the Poisson process. This could be due to several factors: Firstly, the Hawkes process introduces more complex dynamics and higher volatility, which can lead to greater instability in market order flow. This may pose significant challenges for effective risk management by the market maker, thereby impacting overall performance. Secondly, differences in process modeling and algorithm implementation might contribute to performance discrepancies, with the Hawkes process potentially being less efficient or stable compared to the Poisson process. Moreover, the Hawkes process may require different parameter adjustments and strategy optimisations to account for its unique market dynamics, affecting the market maker's wealth mean and Sharpe ratio.

Figure \ref{figure_ALL} illustrates the performance of five types of MM agents at $\eta=0.0$ and $\zeta=0.0$. Each gray plane in the figure represents a 2D plot where six curves depict the return distributions for different agents under different adversarial conditions. The x-axis corresponds to the terminal return values, while the y-axis indicates the density of these distributions. The color-coded points along each curve represent the Sharpe ratios, with the color bar providing a reference for these ratios. Figure \ref{Always Quoting MM}, \ref{2action}, \ref{4Action_1}, \ref{4Action_2}, and \ref{4Action_3} show the terminal wealth distributions of various market makers under different adversarial environments at $\eta=0.0$ and $\zeta=0.0$.

\vspace{10pt}

\noindent\textbf{2-Action MM (Train \& Test @ vol=2) \& 4-Action MM (Train \& Test @ vol=2).} These two market maker (MM) strategies require the agent to decide at each time step whether to quote or make a unilateral quote. If the decision is to quote, they use the well-trained Always Quoting MM strategy to set the specific prices. Experimental results indicate that, despite having the option to refrain from quoting or to quote unilaterally, the agents choose to provide both ask and bid prices over 90\% of the time. This behavior aligns with real-world market-making requirements, such as the London Stock Exchange’s rule \cite{LSEDM} that mandates registered market makers to maintain quotes for at least 90\% of continuous trading during mandatory periods, or the Deutsche Börse Cash Market’s MiFID II compliance \cite{mifid2}, which requires regulated market makers to quote for at least 50\% of daily trading hours on a monthly average. Additionally, both 2-Action MM (Train \& Test @ vol=2) and 4-Action MM (Train \& Test @ vol=2) strategies show a slight improvement in Sharpe ratio compared to the Always Quoting MM strategy.

Several factors may contribute to this phenomenon. First, the Always Quoting MM (Train \& Test @ vol=2) agent, trained through adversarial reinforcement learning, has developed the ability to control risk effectively through appropriate quoting strategies. Therefore, even when given more options, the agent tends to stick with this well-trained risk management approach. Second, the simulated market environment in the experiment may lack sufficient volatility. Even if volatility is present, the risk per trade remains relatively limited due to volume constraints.

Moreover, in the Hawkes process, the probability of order matching is significantly influenced by the outcome of the previous trade. A completed trade increases the probability of at least one subsequent trade, directly affecting the market maker’s quoting strategy. Specifically, in the experimental setup, the Hawkes process may exhibit a "rapid intensity increase and slow intensity decay" characteristic. After each successful trade, the intensity in the Hawkes process rises significantly, leading to a higher probability of subsequent trades in the short term. The slow decay of intensity means that the market maintains a high level of trading activity shortly after a transaction. This mechanism in the Hawkes process encourages market makers to quote more actively. Even in low liquidity environments, market makers are more willing to quote prices that facilitate trades, thereby enhancing liquidity. In low liquidity markets, market makers might opt for ensuring trade execution over immediate profit, which contributes to overall market liquidity. This strategy not only aids the market makers' survival and development but also helps improve the overall liquidity and stability of the market.

\vspace{10pt}

\noindent\textbf{4-Action MM (Train @ vol=2, Test @ vol=200) \& 4-Action MM (Train \& Test @ vol=200).} Volatility, as a key market environment indicator, can prompt market makers to adjust their trading strategies. This paper investigates how increasing volatility from 2 to 200 significantly amplifies price changes per time step, and evaluates the performance of the 4-Action MM strategy trained at Volatility=2 when tested at Volatility=200. The results indicate that, compared to 4-Action MM (Train \& Test @ vol=2), the 4-Action MM (Train @ vol=2, Test @ vol=200) strategy maintains relatively stable performance in high-volatility environments. In contrast, the 4-Action MM (Train \& Test @ vol=200) strategy often opts to refuse quotes or make unilateral quotes, resulting in a Sharpe ratio significantly lower than that of the former. The differences in performance between these strategies under varying volatility conditions may stem from their adaptation to training and testing environments. 

For the 4-Action MM (Train @ vol=2, Test @ vol=200) strategy: Trained in a low-volatility (Volatility=2) environment, the agent becomes accustomed to stable market conditions. The adversarial agents and Always Quoting MM strategy used during training were also based on this low-volatility setting, allowing the agent to learn effective quoting and risk management techniques for such conditions. When tested in a high-volatility environment (Volatility=200), the agent leverages its learned general strategies from the low-volatility training to adapt. This adaptability enables the agent to maintain relatively stable performance in the new high-volatility environment.

For the 4-Action MM (Train \& Test @ vol=200) strategy: Trained in a high-volatility (Volatility=200) environment, the agent encounters extreme market fluctuations. This training likely results in the agent learning strategies specifically tailored for high-volatility markets, such as more cautious bilateral quoting or frequent strategy adjustments. However, because the adversarial agents and Always Quoting MM strategy used during training were based on a low-volatility environment, the agent may face misalignment with quoting strategies under high volatility. This mismatch may lead the agent to adopt more conservative approaches, such as refusing to quote or making unilateral quotes, to mitigate the risks associated with extreme market fluctuations, consequently lowering its Sharpe ratio.

\section{Conclusions} 
This study highlights the effectiveness of integrating Adversarial Reinforcement Learning (ARL) with sophisticated modeling techniques such as Hawkes Processes and variable volatility levels to advance market-making strategies. By expanding the action space and employing a more realistic representation of market dynamics, our research demonstrates that market makers show significant resilience and adaptability across different volatility regimes. The 4-Action MM (Train @ vol=2, Test @ vol=200) performs stably in high-volatility environments due to its training strategy's strong adaptability and robustness. In contrast, the 4-Action MM (Train \& Test @ vol=200) performs poorly in high-volatility environments, primarily because the always-quoting strategy used during training was generated under a volatility of 2, leading to a mismatch with the testing environment. This mismatch causes the agent to adopt more conservative quoting strategies. 

Therefore, future work could explore whether market makers in high-volatility environments would still choose to refrain from quoting or use unilateral quoting to mitigate risks if they had the capability to earn through bid-ask spreads. Specifically, it would be interesting to train a 4-Action MM under a volatility of 200 and examine whether it chooses to quote and, if so, provides specific prices using strategies trained under this high volatility.

\bibliographystyle{ACM-Reference-Format}
\bibliography{main}

\end{document}